\documentstyle[11pt,paspconf]{article}
\def\msun{$ {\rm M}_\odot \ $}
\def\msunp{$ {\rm M}_\odot$}
\def\HII{H\,{\sc ii} }

\def\I{{\'\i}}
\def\kms{${\rm km~s}^{-1}$}

\begin{document}

\title{Some Implications of Galactic Abundance Gradients Derived From
\HII Regions and Planetary Nebulae}
\author{Manuel Peimbert and Leticia Carigi}
\affil{Instituto de Astronom\I a-UNAM,\\ Apdo Postal 70-264, M\'exico
D. F., M\'exico}

\begin{abstract}
By comparing observed abundance gradients with those predicted by a
chemodynamical model of the Galaxy, the following results are obtained: 
there is a deficiency of O/H poor PNe in the solar vicinity implying
that only a fraction of intermediate mass stars produces PNe; a similar
result is obtained from the observed abundance distributions of 
Ne/H, S/H and Ar/H; the age distributions of the stellar progenitors 
of type II and type III PNe show a substantial overlap, but the 
average age of the progenitors of type III PNe is larger
than that of the progenitors of type II PNe; the S/H and the Ar/H
gradients of type I PNe and \HII regions are similar; the flatter O/H 
gradient of type I PNe relative to that of \HII regions is probably 
due to the effect of temperature fluctuations on the abundance determinations;
the similarity of the abundance gradients derived from \HII regions
and PNe implies that transient phenomena are not important in shaping 
the present day \HII gradients.
\end{abstract}


\section{Introduction}

In trying to fit a chemodynamical model of the galaxy to the observed
radial abundance gradients derived from \HII regions and planetary nebulae
many assumptions have to be made. In this note, we discuss the effect
on the assumptions of the comparison of the chemodynamical model by 
Allen et al. (1998) with the observations of \HII regions by Peimbert
(1979), Shaver et al. (1983), and V\I lchez \& Esteban (1996)
and of planetary nebulae by Maciel \& K\"oppen (1994).  \HII regions
give us information on the present value of the abundance distributions
in the interstellar medium (ISM), while PNe give us information on the abundance
distributions in the ISM at the time of the formation of the progenitor 
stars of those elements not affected by stellar evolution.  
Maciel \& K\"oppen have used the classification by Peimbert (1978), where
type I PNe come from the high mass end of the intermediate mass stars (IMS) 
(those in the 0.83 \msun to 8.4 \msun range), and types II and III from the 
intermediate and low mass end of the IMS, respectively.

\begin{figure}
\plotfiddle{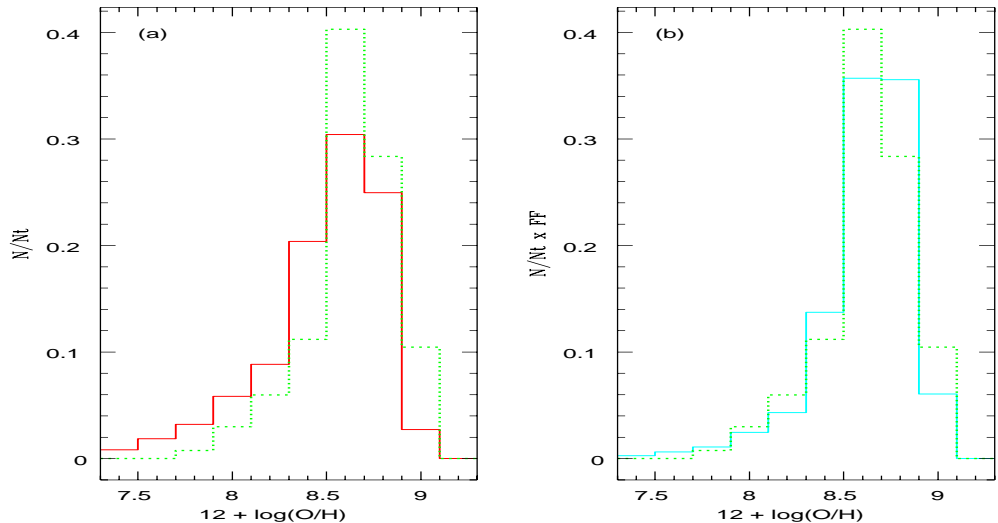}{65truemm}{0}{70}{37}{-210}{-65}
\caption{The solid lines represent the model (Allen et al. 1998) and the
dotted lines the observations (Maciel \& K\"oppen 1994). Panel (a) shows the
PNe O/H distribution assuming that all IMS produce PNe; panel (b) presents
also the PNe O/H distribution assuming that only the fraction of IMS given
by Allen et al. produces PNe.}
\end{figure}

\section{The Planetary Nebula Problem}

In Figure 1a, we present the O/H distribution for all the
observed PNe as predicted by the
model by Allen et al. (1998) under the assumption that all IMS produce 
PNe. This is compared with  the
observed O/H distribution by Maciel \& K\"oppen (1994). The observed 
distribution shows a defficiency of O/H poor objects relative to the 
predicted distribution. A similar result is obtained for the Ne/H, S/H 
and Ar/H distributions, ruling out the possibility that O/H poor objects 
increase substantially their O/H ratio during their evolution.
We call this deficiency of O/H poor objects the PN problem, in
analogy with the G dwarf problem. It should be noted that the chemodynamical
model explains the G dwarf problem and that a different solution has to 
be found for the PN problem. 

Allen et al. (1998) suggest that only about half of the IMS produce  PNe, 
moreover they proposed a fraction function, FF, that gives the probability
that a star of a given mass will produce a PN. In Figure 1b,
we present the expected O/H distribution adopting the FF proposed
by Allen et al. There are other results discussed by Allen et al. 
that support the idea that only a fraction of IMS produce PNe: a) the
scale height of PNe in the solar neighborhood, b) the estimated PNe
birth rate for the solar neighborhood, c) the decrease of the PNe birthrate
with M$_{\rm bol}$ and (B-V)$_0$ in extragalactic systems, and d) the observed fraction 
of white dwarfs with masses in the 0.4 \msun to 0.55 \msun range that presumably
did not go through the PN stage. 

In Figure 2, we present the O/H cumulative function for PNe; from this
figure it can also be noted that without the FF there is a defficiency
of O/H poor PNe.

\begin{figure}
\plotfiddle{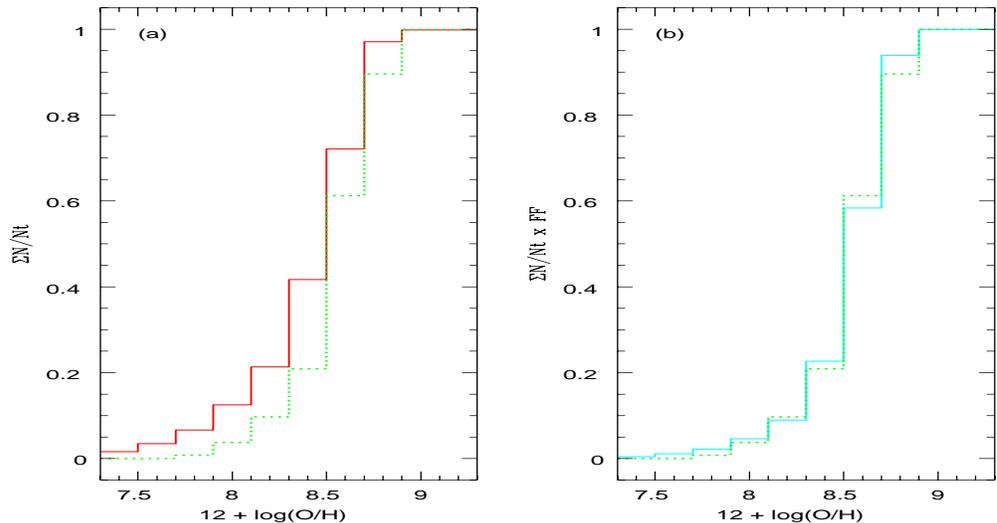}{65truemm}{0}{70}{37}{-210}{-65}
\caption{The solid lines represent the model (Allen et al. 1998) and the
dotted lines the observations (Maciel \& K\"oppen 1994). Panel (a) shows the
PNe O/H cumulative function assuming that all IMS produce PNe; panel (b) presents
also the PNe O/H cumulative function assuming that only the fraction of
IMS given by Allen et al. produce PNe.}
\end{figure}

\section{Planetary Nebula Types}

Peimbert (1978, 1990) divided PNe in four types according to the mass of the
progenitor in the main sequence. To estimate the mass of the progenitor of 
a given PN , and consequently its type, one can use its chemical 
composition or its dynamical properties. From models, it is trivial to divide
PNe among different types; from observations it is difficult to estimate the
mass of the progenitor stars of PNe.

Allen et al. (1998) defined those PNe with peculiar velocities higher than
60 \kms\ as type III, and those with peculiar velocities lower than 60 \kms
\ as type II. From Allen et al., it follows that there is an
overlap in initial masses between PNe of types II and III. 
The average mass of the progenitors of type II PNe is  
1.29 \msun, corresponding to an average age of 3.6 Gyr, while for type III PNe, 
the averages correspond to 1.05 \msun and 6.4 Gyr respectively. This division
is in good agreement with the fractions of the observed PNe by 
Maciel \& K\"oppen (1994), that amount to 39, 45, and 16 percent for types I, II, 
and III, respectively. We have adopted a solar Galactocentric distance  
of 8 kpc and have considered only those PNe with Galactocentric distances in the 
6 to 10 kpc range. We have not estimated the incompleteness effect on the 
sample by Maciel and K\"oppen due to dust absorption in the plane of the Galaxy; 
this effect goes in the opposite direction to that needed to solve the PN 
problem, since the incompleteness is higher for type I and II than for 
type III PNe.

We can also divide type II and III PNe strictly by mass, adopting the
FF of Allen et al. (1998) and the relative fractions of
observed types by Maciel \& K\"oppen (1994). The limiting mass between
both types becomes 0.97 \msunp. The following
averages are obtained from the chemodynamical model by Allen et al. (1998): 
$12 + \log (\rm O/H)$ = 8.62, and M = 1.25 \msun for type II, while $12 + 
\log (\rm O/H)$ = 8.20, 
and M = 0.92 \msun for type III. The average masses correspond to 4.0 Gyr for 
type II and to 9.5 Gyr for type III.  As we saw in 
the previous paragraph, it is not possible to have a sharp mass boundary between
types II and III based on dynamical arguments only. Neither is it possible 
to have a sharp mass boundary based on  
the observed abundances for the following reasons: a) there are 
observational errors in the abundance determinations, b) at a given 
Galactocentric distance there are PNe that originated at 
different Galactocentric distances, and c) probably there is a scatter 
in the chemical abundances present in the ISM for a given time and a given 
Galactocentric distance.

Using the set of Maciel \& K\"oppen (1994), we derive average values
of $12 + \log (\rm O/H)$ = 8.66 and 8.35 for types II and III respectively (see Table
1). On the other hand, we can also divide PNe types by abundances only;
therefore by assuming that type III PNe comprise 16\% of the objects with the
lowest O/H values in the set of Maciel \& K\"oppen, and that the rest of non 
type I  are of type II (45\%), we obtain  averages of $12 + \log (\rm O/H)$ = 8.28 and
8.69 for types III and II respectively (see Table 1). The similarity of the
O/H values implies that type II and type III PNe can be divided by O/H
values or by dynamical properties. Also in Table 1, we present the O/H average
values predicted by the model of Allen et al. (1998) for the two types of PNe,
divided by velocities and by abundances, assuming that all the IMS produce PNe
and that only a fraction of the IMS produce PNe. By comparing the observed
values with the predicted ones, it can also be concluded that a better fit to
the observations can be made with the use of the FF  proposed by Allen et al.
Note that in the model predictions based on abundances, the limiting mass between
type I and type II PNe differs from that adopted by Allen et al.

\begin{table}[t]
\caption{Average $12 + \log (\rm O/H)$ values for PN
types divided by velocities and by abundances.
The values in parenthesis correspond to the fraction of
PNe in each category} \label{tbl-1}
\begin{center}
\begin{tabular}{lccc}
\hline
 & Model & Model x FF & Observed \\
\tableline
 &  & & \\
Type II \ ( v $<$ 60 \kms) & 8.48 (41\%) & 8.58 (47\%) & 8.66 (45\%) \\
Type III  ( v $>$ 60 \kms) & 8.28 (44\%) & 8.40 (20\%) & 8.35 (16\%) \\
 &  & & \\
Type II \ ( high O/H ) & 8.46 (45\%) & 8.62 (45\%) & 8.69 (45\%) \\
Type III  (  low O/H ) & 7.89 (16\%) & 8.20 (16\%) & 8.28 (16\%) \\
\hline
\hline
\end{tabular}
\end{center}
\end{table}

\section{Other Implications}
The O/H and Ne/H gradients derived from type I PNe are considerably
smaller than those derived from \HII regions. On the other hand, the
S/H and the Ar/H gradients are similar. The difference in the O/H 
gradients could be due to ON cycling or to the presence of temperature
fluctuations. The presence of ON cycling in PNe is controversial
(e.g. Torres-Peimbert \& Peimbert 1997, and references therein);
Peimbert et al. (1995) argue that the low O/H values derived for 
type I PNe are  due to the assumption of a constant temperature in 
the abundance determinations; taking temperature fluctuations into account,
they find that ON cycling is not present in type I PNe. 
The presence of temperature fluctuations affects more 
the Ne and O than the S and Ar abundance determinations, when these
abundances are derived from nebular lines. This is because the excitation energies of
the O and Ne lines are higher than those of the S and Ar lines. Moreover it
can  be shown that the effect goes in the direction of flattening
the gradients, since the higher the heavy element abundances, the lower
the temperature and the higher is the effect of temperature fluctuations 
on the abundance determinations.

The similarity of the abundance gradients derived from \HII regions, PNe
of type I (S and Ar), PNe of type II, and PNe of type III implies that
transient phenomena are not important in shaping the present day \HII
gradients. PNe seem to show a  flattening of the abundance
gradients at large Galactocentric distances similar to that
of \HII regions (Maciel 1997, private communication), these
flattenings are not explained by the model of Allen et al. (1998,
see also Carigi 1996), and probably imply that gas flows are 
important in the outer regions of the Galaxy (see reviews by 
Peimbert 1995; Maciel 1997).



\end{document}